# A SEARCH FOR INTEGRABLE FOUR-DIMENSIONAL NONLINEAR ACCELERATOR LATTICES*

S. Nagaitsev# FNAL, Batavia, IL 60510, U.S.A
V. Danilov SNS, Oak Ridge, TN 37830, U.S.A.

*Abstract*

Integrable nonlinear motion in accelerators has the potential to introduce a large betatron tune spread to suppress instabilities and to mitigate the effects of space charge and magnetic field errors. To create such an accelerator lattice one has to find magnetic and/or electric field combinations leading to a stable integrable motion. This paper presents families of lattices with one invariant where bounded motion can be easily created in large volumes of the phase space. In addition, it presents two examples of integrable nonlinear accelerator lattices, realizable with longitudinal-coordinate-dependent magnetic or electric fields with the stable nonlinear motion, which can be solved in terms of separable variables.

## INTRODUCTION

All present accelerators (and storage rings) are built to have "linear" focusing optics (also called lattice). The lattice design incorporates dipole magnets to bend particle trajectory and quadrupoles to keep particles stable around the reference orbit. These are "linear" elements because the transverse force is proportional to the particle displacement, $x$ and $y$. This linearity results (after the action-phase variable transformation) in a Hamiltonian of the following type:

$$H(J_1, J_2) = \nu_x J_1 + \nu_y J_2, \quad (1)$$

where $\nu_x$ and $\nu_y$ are betatron tunes and $J_1$ and $J_2$ are actions. This is an integrable Hamiltonian. The drawback of this Hamiltonian is that the betatron tunes are constant for all particles regardless of their action values. It has been known since early 1960-s that the spread of betatron tunes is extremely beneficial for beam stability due to the so-called Landau damping. However, because the Hamiltonian (1) is linear, any attempt to add non-linear elements (sextupoles, octupoles) to the accelerator generally results in a reduction of its dynamic aperture, resonant behavior and particle loss. We propose [1] that the accelerator lattice can be designed to be integrable and, at the same time, non-linear, such that the betatron tunes would depend on actions $J_1$ and $J_2$, thus creating a natural spread in betatron tunes. We expect that for certain classes of such integrable non-linear Hamiltonians there occurs a non-linear stabilization of resonances in a manner similar to a resonantly perturbed pendulum [2].

___________________________________________
*Work supported by UT-Battelle, LLC and by FRA, LLC for the U. S. DOE under contracts No. DE-AC05-00OR22725 and DE-AC02-07CH11359 respectively.
#nsergei@fnal.gov

## SPECIAL NON-LINEAR FIELDS

Let us assume that we have equal linear focusing in the horizontal and vertical planes and some additional time-dependent potential. The Hamiltonian has the form

$$H = \frac{p_x^2}{2} + \frac{p_y^2}{2} + K(s)\left(\frac{x^2}{2} + \frac{y^2}{2}\right) + V(x,y,s), \quad (2)$$

where $K(s)$ is the linear focusing coefficient and $s$ is the time-equivalent longitudinal coordinate. Now we will make a normalized-variable substitution to obtain the following Hamiltonian:

$$H_N = \frac{p_{xN}^2 + p_{yN}^2}{2} + \frac{x_N^2 + y_N^2}{2} + U(x_N, y_N, \psi), \quad (3)$$

where

$$U(x_N, y_N, \psi) = \beta(\psi)V\left(x_N\sqrt{\beta(\psi)}, y_N\sqrt{\beta(\psi)}, s(\psi)\right) \quad (4)$$

and $\psi$ is the "new time" variable defined as the betatron phase,

$$\psi' = \frac{1}{\beta(s)}. \quad (5)$$

Three main ideas of the present proposal are as follows:

1. The potential $U$ in equation (4) can be chosen such that it is time-independent (see next section for examples). This results in time-independent Hamiltonian (3).
2. Among all such chosen time-independent Hamiltonians we can find several sets of potentials $U$, which obey the Laplace equation and posses the second integral of motion – such systems are exactly integrable and realizable by magnetic or electric fields!
3. The equal horizontal and vertical focusing can be obtained by solenoids or by linear elements (quads and dipoles) to the extent that such elements result in a thin-lens transfer matrix.

Consider an element of lattice periodicity consisting of two parts: (1) a drift space, $L$, with exactly equal horizontal and vertical beta-functions, followed by (2) an optics insert, $T$, which has the transfer matrix of a thin axially symmetric lens (Figure 1).

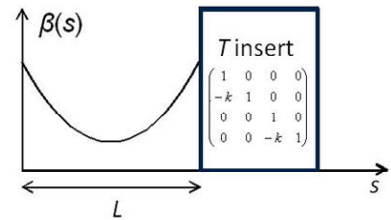

Figure 1: An element of periodicity: a drift space with equal beta-functions followed by a $T$ insert.

An example of beam-line optics to realize such an element of periodicity is shown in Figure 2.

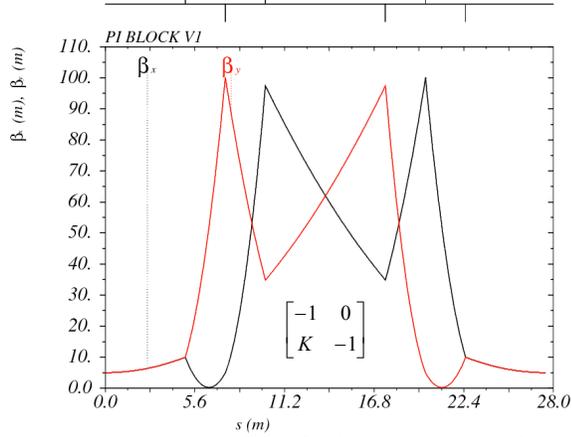

Figure 2: An example of linear optics with six quadrupoles resulting in equal beta-functions and equal transfer matrices.

The betatron phase advance per this element of periodicity can vary from $\pi$ to $2\pi$ depending on the value of $K$ and the drift space length $L$.

## AN EXAMPLE WITH OCTUPOLES

Let us consider some large number (>20) of short octupoles occupying the entire drift section $L$ in Fig. (1). In order to obtain the time-independent Hamiltonian it is clear that the strength of these octupoles along the drift section needs to scale as $\beta(s)^{-3}$:

$$V(x,y,s) = \frac{\kappa}{\beta(s)^3}\left(\frac{x^4}{4} + \frac{y^4}{4} - \frac{3x^2 y^2}{2}\right). \quad (6)$$

This results in

$$U = \kappa\left(\frac{x_N^4}{4} + \frac{y_N^4}{4} - \frac{3y_N^2 x_N^2}{2}\right), \quad (7)$$

where $\kappa$ is an arbitrary constant (the desired octupole strength).

The resulting Hamiltonian with such a potential is non-integrable but this example could be useful in obtaining a modest (10%) betatron tune spread. The resulting potential energy, which is the sum of a harmonic oscillator potential and the potential $U$, is shown in Fig. 3.

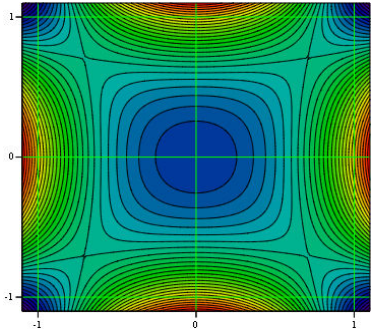

Figure 3: A contour plot of the potential energy for a harmonic oscillator with an octupole potential.

From Fig. 3 one can notice that there is a maximum value of total energy, $= (4\kappa)^{-1}$, above which the trajectories are unbound. This limits the maximum attainable tune spread to about 12% of the linear tune (independent of the octupole strength $\kappa$).

## INTEGRABLE POTENTIALS

Let us consider now two non-linear integrable potentials.

### Case 1

From Eq. (4) one can notice that if the potential $V(x,y,s)$ scales as $z^{-2}$ ($z$ is $x$ or $y$), then the time-dependence of the potential $U$ naturally disappears. An example of such a potential is given by Eq. (8):

$$V(x,y) = U(x,y) = \frac{a(x^2 - y^2) + 2bxy}{(x^2 + y^2)^2}, \quad (8)$$

where $a$ and $b$ are arbitrary coefficients.

This potential satisfies the Laplace equation and thus can be implemented in practice using magnetic or electric fields in vacuum. The Hamiltonian (4) (from now on we will drop the subscript $N$) with such a potential has two integrals of motions: the total energy, $E = H$, and the second integral,

$$I = (xp_y - yp_x)^2 + 2\frac{a(x^2 - y^2) + 2bxy}{x^2 + y^2}. \quad (9)$$

These are the non-linear analogs of the Courant-Snyder invariants.

The potential (8) has a singularity at the origin, but not all trajectories encircle it. One can prove that if

$$I > 2\sqrt{a^2 + b^2} \quad (10)$$

the trajectory never encircles the singularity. This condition can be satisfied at injection into the ring by choosing the appropriate initial beam distribution.

The lowest-order multipole expansion (away from the singularity) of such a potential is a dipole. Thus, this potential can be used for bending particle trajectories, while also providing nonlinear focusing.

### Case 2

G. Darboux [3] has conducted the first systematic study of integrable potentials with the second integral of motion quadratic in momentum. He has demonstrated that such potentials can be presented in elliptic coordinates in the following way

$$U(x,y) = \frac{f(\xi) + g(\eta)}{\xi^2 - \eta^2}, \quad (11)$$

where $f$ and $g$ are arbitrary functions,

$$\xi = \frac{\sqrt{(x+c)^2 + y^2} + \sqrt{(x-c)^2 + y^2}}{2c}$$
$$\eta = \frac{\sqrt{(x+c)^2 + y^2} - \sqrt{(x-c)^2 + y^2}}{2c} \quad (12)$$

are elliptic variables and $c$ is an arbitrary constant.

The second integral of motion yields

$$I(x,y,p_x,p_y) = (xp_y - yp_x)^2 + c^2 p_x^2 + 2c^2 \frac{f(\xi)\eta^2 + g(\eta)\xi^2}{\xi^2 - \eta^2} \quad (13)$$

First, we would notice that the harmonic oscillator potential $(x^2 + y^2)$ can be presented in the form of Eq. (11) with $f_1(\xi) = c^2\xi^2(\xi^2 - 1)$ and $g_1(\eta) = c^2\eta^2(1-\eta^2)$. Second, we have found the following family of potentials that satisfy the Laplace equation and, at the same time, can be presented in the form of Eq. (11):

$$f_2(\xi) = \xi\sqrt{\xi^2 - 1}(d + t\,\mathrm{acosh}(\xi)),$$
$$g_2(\eta) = \eta\sqrt{1-\eta^2}(q + t\,\mathrm{acos}(\eta)) \quad (14)$$

where $d$, $q$, and $t$ are arbitrary constants. Thus, the total potential energy is given by

$$U(x,y) = \frac{x^2}{2} + \frac{y^2}{2} + \frac{f_2(\xi) + g_2(\eta)}{\xi^2 - \eta^2}. \quad (15)$$

Of a particular interest is the potential with $d = 0$ and $q = -\frac{\pi}{2}t$, because its lowest multipole expansion term is a quadrupole. Figure 4 presents a contour plot of the potential energy (15) for $c = 1$ and $t = -0.4$.

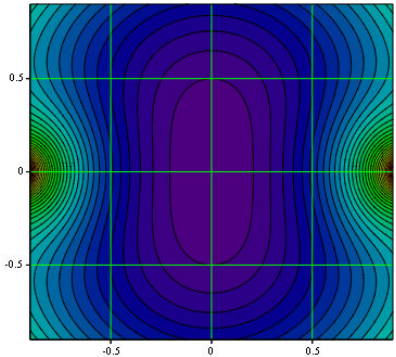

Figure 4: A contour plot of the potential energy Eq. (15) with $c = 1$ and $t = -0.4$.

The multipole expansion of this potential for $c = 1$ is as follows:

$$U(x,y) \approx \frac{x^2}{2} + \frac{y^2}{2}$$
$$-t\,\mathrm{Re}\left((x+iy)^2 + \frac{2}{3}(x+iy)^4 + \frac{8}{15}(x+iy)^6 + \frac{16}{35}(x+iy)^8 + ...\right) \quad (16)$$

Let us now determine the maximum attainable betatron frequency spread in such a potential. First, this potential provides additional focusing in $x$ for $t < 0$ and defocusing in $y$. Thus, for a small-amplitude motion to be stable, one needs $-0.5 < t < 0$. This corresponds to the following small-amplitude betatron frequencies,

$$\nu_x = \nu_0\sqrt{1-2t}$$
$$\nu_y = \nu_0\sqrt{1+2t}, \quad (17)$$

where $\nu_0$ is the unperturbed linear-motion betatron frequency. For large amplitudes both betatron frequencies approach $\nu_0$. Thus, a spread of 100% in one plane and ~40% in another is possible in an ideal lattice with thin lenses. If one uses a linear optics insert with a phase advance of $\pi$, similar to Figure 2, the maximum tune spread would be halved, 50% and 20%. Figure 5 shows an example of the trajectory (in normalized coordinates) for $t = -0.2$.

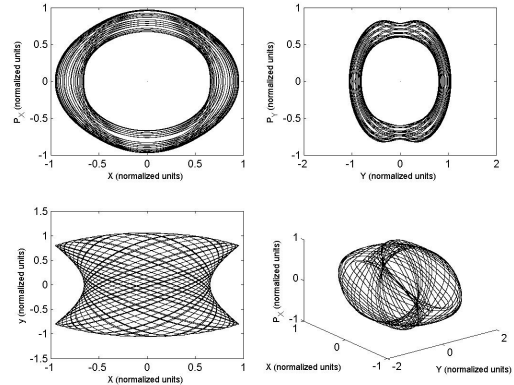

Figure 5: Two-dimensional projections of integrable motion: horizontal phase space (left top), vertical phase space (right top), and x-y projection (left bottom) and the 3-dimensional trajectory surface in px, x, y coordinates.

To realize this potential in practice one could install a number (20-50) of short nonlinear elements such that the time dependence in Eq. (4) is cancelled approximately. Such a nonlinear lattice would have two integrals of motion and a large betatron frequency spread.

## SUMMARY

This paper presents two exactly integrable nonlinear accelerator lattices realizable with magnetic or electric fields. The phase space occupied by trajectories has large regions of stability and a very large betatron frequency spread that can mitigate instabilities, space charge effects, and particle loss. In addition, it demonstrates that there exists a variety of nonlinear 2D lattices with one integral of motion which can be used for the creation of a bound (but chaotic) motion or many other integrable cases. Possible practical lattice constructions are discussed.

## ACKNOWLEDGEMENTS

We would like to thank Alexander Valishev (FNAL) for many fruitful discussions.